\documentclass{article}

    \usepackage[preprint]{neurips_2025}

\usepackage[utf8]{inputenc} 
\usepackage[T1]{fontenc}    
\usepackage{hyperref}       
\usepackage{url}            
\usepackage{booktabs}       
\usepackage{amsfonts}       
\usepackage{nicefrac}       
\usepackage{microtype}      
\usepackage{xcolor}         
\usepackage{natbib}
\usepackage{algorithm}
\usepackage{amsthm}
\usepackage{amsmath}
\usepackage{algpseudocode}
\usepackage{placeins}
\usepackage{multirow} 
\usepackage{colortbl}
\usepackage{boldline}
\usepackage{graphicx}
\theoremstyle{definition}

\newtheorem{theorem}{Theorem}

\newtheorem{remark}{Remark}

\title{Toward Fair Federated Learning under Demographic Disparities and Data Imbalance}

\author{%
  Qimimg Wu\thanks{These authors contributed equally to this work.} \\
  Duke-NUS Medical School \\
  \texttt{wuqiming@u.nus.edu} \\
  \And
  Siqi Li\footnotemark[1] \\
  Duke-NUS Medical School \\
  \texttt{siqili@u.duke.nus.edu} \\
  \AND
  Doudou Zhou\\
  National University of Singapore \\
  \texttt{ddzhou@nus.edu.sg} \\
  \And
  Nan Liu\thanks{Correspondence: Nan Liu, Centre for Quantitative Medicine, Duke-NUS Medical School, 8 College Road, Singapore 169857, Singapore. Email: \texttt{liu.nan@duke-nus.edu.sg}} \\
  Duke-NUS Medical School \\
  \texttt{liu.nan@duke-nus.edu.sg} \\
}

\begin{document}

\maketitle

\begin{abstract}
Ensuring fairness is critical when applying artificial intelligence to high-stakes domains such as healthcare, where predictive models trained on imbalanced and demographically skewed data risk exacerbating existing disparities. Federated learning (FL) enables privacy-preserving collaboration across institutions, but remains vulnerable to both algorithmic bias and subgroup imbalance—particularly when multiple sensitive attributes intersect.
We propose \textbf{FedIDA} (\underline{Fed}erated Learning for \underline{I}mbalance and \underline{D}isparity \underline{A}wareness), a framework-agnostic method that combines fairness-aware regularization with group-conditional oversampling. FedIDA supports multiple sensitive attributes and heterogeneous data distributions without altering the convergence behavior of the underlying FL algorithm.
We provide theoretical analysis establishing fairness improvement bounds using Lipschitz continuity and concentration inequalities, and show that FedIDA reduces the variance of fairness metrics across test sets. Empirical results on both benchmark and real-world clinical datasets confirm that FedIDA consistently improves fairness while maintaining competitive predictive performance, demonstrating its effectiveness for equitable and privacy-preserving modeling in healthcare. The source code is available on \href{https://github.com/qimingwu/FedIDA}{GitHub}.
\end{abstract}

\section{Introduction}
As machine learning (ML) systems are increasingly deployed in high-stakes domains such as healthcare and social science, concerns about fairness and equity have become central to both research and practice\cite{Caton_2024}. Predictive models that perform unequally across demographic subgroups—defined by attributes such as gender, race, or socioeconomic status—can reinforce or even exacerbate existing disparities. Federated learning (FL), which enables model training across multiple institutions without sharing raw data, offers a promising solution for preserving privacy in distributed environments. However, FL brings new challenges for fairness due to non-IID data distributions, limited representation of minority groups, and heterogeneity across participating sites.

Existing fairness-aware FL methods often suffer from two key limitations. First, most frameworks assume a single sensitive attribute and do not generalize to real-world settings where multiple intersecting sensitive variables are present. Second, these methods rarely address the joint impact of algorithmic disparities and data imbalance—particularly the imbalance across sensitive subgroups—which can be further exacerbated in federated settings due to fragmentation and skewed local distributions.

To address these gaps, we propose a framework-agnostic solution, \textbf{\underline{Fed}}erated Learning for \textbf{\underline{I}}mbalance and \textbf{\underline{D}}isparitiy \textbf{\underline{A}}wareness (\textbf{FedIDA}),
that improves fairness in FL by jointly addressing both algorithmic bias and data imbalance. Our approach introduces a group-wise fairness regularization that supports multiple sensitive attributes, combined with a novel fairness-aware oversampling procedure to balance underrepresented (sensitive attributes, outcome) pairs at the local client level. The resulting formulation is convex, theoretically well-behaved, and compatible with any standard FL optimizer.

We provide theoretical analysis showing that FedIDA improves fairness stability via subgroup-balanced oversampling, with variance bounds derived using Lipschitz continuity and concentration inequalities. Experiments on a large-scale, real-world clinical dataset demonstrate that our method substantially reduces disparities across sensitive groups without sacrificing predictive performance, under both balanced and imbalanced conditions. In addition to empirical evaluation, we theoretically characterize how FedIDA improves the stability of fairness metrics by reducing their variance under test-time sampling, leveraging concentration inequalities and Lipschitz properties.

\section{Related Work}

\paragraph{Algorithmic Disparities in FL.}
Several studies have investigated fairness in the context of FL, particularly by modifying objective functions or applying reweighting techniques to mitigate algorithmic disparities. However, most of these methods focus on a \textit{single sensitive attribute} (e.g., gender or race)~\cite{cui2021addressing, du2020fairnessawareagnosticfederatedlearning, abay2020mitigatingbiasfederatedlearning, 9378043}, which limits their applicability in real-world settings where fairness concerns typically arise from the intersection of multiple demographic factors. Our work addresses this limitation by introducing a convex fairness penalty that accommodates \textit{multiple sensitive attributes}, enabling subgroup-aware learning in a more realistic and generalizable way.

\paragraph{Data Imbalance in FL.}
FL is also vulnerable to data imbalance, including both class imbalance (e.g., skewed outcome prevalence) and subgroup imbalance (e.g., under-representation of specific sensitive subpopulations). While prior work has explored reweighting or resampling techniques to address class imbalance~\cite{hong2024trans, wang2021addressing, 9616052}, few methods explicitly consider the interaction between imbalance and fairness—particularly when subgroup sizes are small and unevenly distributed across clients. Our method directly tackles this issue through a fairness-aware oversampling procedure that balances sensitive-outcome pairs locally.

\paragraph{Joint Fairness and Imbalance.}
To the best of our knowledge, few existing methods jointly address both algorithmic disparities and subgroup-level imbalance in FL settings. Our proposed framework fills this gap by combining fairness regularization with group-aware oversampling in a manner that is compatible with any FL optimizer. This enables improved fairness across multiple sensitive attributes under heterogeneous and imbalanced data distributions, without sacrificing predictive performance.

\section{Method}
\subsection{Preliminaries}

We consider a standard supervised classification problem, where the objective is to predict a binary outcome 
$
\mathbf{Y} \in \mathcal{Y} = [-1,1]
$
from an input feature set 
$
\mathbf{X} \in \mathcal{X}= \mathbb{R}^{p_A+p_S},
$
with the total feature dimension given by $p = p_A + p_S$. We assume that the feature set can be decomposed as
$
\mathbf{X} = \{\mathbf{A}, \mathbf{S}\},
$
where $\mathbf{A} \in \mathbb{R}^{p_A}$ represents the non-sensitive features (e.g., clinical measurements) and $\mathbf{S} \in \mathbb{R}^{p_S}$ contains the sensitive attributes (e.g., gender, ethnicity, income). In many ethically critical applications—such as predicting patient treatments—the final decision-making model should rely solely on $\mathbf{A}$ for ethical considerations.

Let $\mathcal{P}$ denote the joint distribution over $\mathcal{X} \times \mathcal{Y}$. 
Let $D = \{(\mathbf{x_i}, y_i)^n_{i=1}\}$ be a training set of $n$ samples drawn i.i.d from $\mathcal{P}$, separated by a total of $K$ groups. Denote these groups by $D_1, D_2, \dots, D_K$, the formation of which depends only on sensitive attributes $\mathbf{S}$, with each group $D_k$ containing $n_k$ samples such that
$\sum_{k=1}^{K} n_k = n$.

We define the overall training objective with a fairness regularizer as
\begin{equation}
L(\mathbf{w}, D) + \lambda f(\mathbf{w}, D),
\end{equation}
where $\mathbf{w} \in \mathbb{R}^{p_A}$ denotes the model parameters, $L(\mathbf{w}, D)$ is the standard loss function (e.g., cross-entropy for classification), and $\lambda$ is the regularization parameter governing the fairness penalty. 

Consistent with prior work~\cite{berk2017convexframeworkfairregression}, we define the group fairness penalty as
\begin{equation}\label{penalty}
f(\mathbf{w}, D) = \frac{1}{\sum_{k \neq k'} n_k n_{k'}} \sum_{\substack{k \neq k' \\ (\mathbf{x}_i, y_i) \in D_k \\ (\mathbf{x}_j, y_j) \in D_{k'}}} 
d(y_i, y_j) \left( \mathbf{w}^\top \mathbf{a}_i - \mathbf{w}^\top \mathbf{a}_j \right)
\end{equation}
where 
$$
d(y_i, y_j) = \mathbf{1}[y_i = y_j]
$$ 
serves as the cross-group fairness weight. This formulation generalizes the conventional binary partition setting to accommodate multiple sensitive groups.

\paragraph{Algorithmic Disparities.}
Algorithmic disparities arise when a predictive model exhibits systematically different behavior across groups defined by sensitive attributes. A widely used criterion is demographic parity (DP), which requires the rate of positive predictions to be invariant across sensitive groups. Formally, a model satisfies DP if
\[
\Pr(f(\mathbf{A}; \theta) = 1 \mid (\mathbf{x}, y) \in D_k) = \Pr(f(\mathbf{A}; \theta) = 1) \quad \forall k \in \{1, \dots, K\}.
\]
Violations of DP are commonly quantified using Demographic Parity Difference (DPD) and Demographic Parity Ratio (DPR):
\[
\text{DPD} = \max_{k, k'} \left| \mathbb{E}[f(\mathbf{A}; \theta) \mid (\mathbf{x}, y) \in D_k] - \mathbb{E}[f(\mathbf{A}; \theta) \mid (\mathbf{x}, y) \in D_{k'}] \right|
\]
\[
\text{DPR} = \frac{\min_{k} \mathbb{E}[f(\mathbf{A}; \theta) \mid (\mathbf{x}, y) \in D_k]}{\max_{k'} \mathbb{E}[f(\mathbf{A}; \theta) \mid (\mathbf{x}, y) \in D_{k'}]}.
\]

This group-based framework extends naturally to error-based metrics such as false positive rate (FPR) and positive predictive value (PPV). We define the following disparity metrics:
\[
\text{DFPR} = \max_{k, k'} \left| \mathbb{E}[f(\mathbf{A}; \theta) \mid Y = 0, (\mathbf{x}, y) \in D_k] - \mathbb{E}[f(\mathbf{A}; \theta) \mid Y = 0, (\mathbf{x}, y) \in D_{k'}] \right|
\]
\[
\text{DPPV} = \max_{k, k'} \left| \mathbb{E}[Y \mid f(\mathbf{A}; \theta) = 1, (\mathbf{x}, y) \in D_k] - \mathbb{E}[Y \mid f(\mathbf{A}; \theta) = 1, (\mathbf{x}, y) \in D_{k'}] \right|.
\]

In this work, we adopt a multi-objective optimization framework that balances predictive performance with group fairness constraints, using these disparity metrics to guide model selection and evaluation.

\paragraph{Data Imbalance.}  
Data imbalance occurs when certain combinations of sensitive attributes and outcome labels are significantly underrepresented in the training data. This skew can lead to models that perform disproportionately well on majority groups while underperforming on minority subgroups, resulting in biased and inequitable predictions. In particular, when a subgroup associated with a specific outcome class is rare, the model may lack sufficient exposure to learn meaningful patterns for that group.

To mitigate this issue, oversampling techniques can be applied to rebalance the training distribution by increasing the representation of underrepresented subgroup–outcome pairs. Methods such as SMOTE~\cite{smote} and ROSE~\cite{rose} can be adapted to generate synthetic data points conditioned on both the sensitive attributes and the target label. These approaches augment the training data in a targeted manner, improving the model’s ability to generalize across all groups and reducing performance disparities arising from data imbalance.

\paragraph{Federated Learning.}  
Federated learning (FL) enables multiple distributed clients to collaboratively train a global model without sharing their raw data. Suppose there are $K$ clients, where the $k$-th client has a local dataset $\mathcal{D}_k$ of size $n_k$. The overall objective is to minimize the empirical risk $\min_{\theta} \; F(\theta) = \sum_{k=1}^{K} \frac{n_k}{n} F_k(\theta)$, with $
F_k(\theta) = \frac{1}{n_k} \sum_{i=1}^{n_k} \ell\Bigl( f(\mathbf{x}_i^{(k)}; \theta), y_i^{(k)} \Bigr)$ and $n = \sum_{k=1}^{K} n_k$. In a typical FL protocol, each client updates its local parameter $\theta_k$ using its own data, and a central server aggregates these updates (e.g., via FedAvg~\cite{mcmahan2023communicationefficientlearningdeepnetworks}):
$\theta \leftarrow \sum_{k=1}^{K} \frac{n_k}{n}\, \theta_k$.

\begin{algorithm}[t]
\caption{\textsc{FairnessAwareROSE}}
\label{FAR}
\begin{algorithmic}[1]
\State \textbf{Input:} Dataset $D = \{(x_i, y_i, s_i)\}_{i=1}^n$, list of feature columns, outcome column, sensitive attribute column, target number of samples per (Sensitive, Outcome) group $N_{\text{target}}$, noise covariance matrix $\Sigma$
\State \textbf{Output:} Balanced dataset $D_{\text{augmented}}$

\State Group dataset $D$ by $(s_i, y_i)$ into subgroups
\State Initialize $D_{\text{augmented}} \gets \emptyset$
\ForAll{subgroup $(s, y)$}
    \State $G \gets$ all examples in subgroup $(s, y)$
    \State $n_{\text{existing}} \gets$ number of examples in $G$
    \If{$n_{\text{existing}} \geq N_{\text{target}}$}
        \State Randomly sample $N_{\text{target}}$ examples from $G$
        \State Add sampled examples to $D_{\text{augmented}}$
    \Else
        \State Add all examples in $G$ to $D_{\text{augmented}}$
        \For{$i = 1$ to $(N_{\text{target}} - n_{\text{existing}})$}
            \State Randomly select $x_0$ from $G$
            \State Sample $\epsilon \sim \mathcal{N}(0, \Sigma)$
            \State $x_{\text{synthetic}} \gets x_0 + \epsilon$
            \State Add $(x_{\text{synthetic}}, y, s)$ to $D_{\text{augmented}}$
        \EndFor
    \EndIf
\EndFor
\State \Return $D_{\text{augmented}}$
\end{algorithmic}
\end{algorithm}
\subsection{FedIDA}

We now present FedIDA (Federated Learning for Imbalance and Disparity Awareness), our proposed algorithm designed to simultaneously address fairness and data imbalance in federated learning (FL) settings. FedIDA extends any standard FL framework $\mathcal{F}$ by incorporating two key components: a fairness penalty term weighted by $\lambda$, and an $\ell_2$ regularization term weighted by $\gamma$. The resulting local objective for a mini-batch $\mathcal{D}_i$ is defined as:

\begin{equation}\label{eq:composite}
F(\mathbf{w}, \mathcal{D}_i) = \mathcal{L}(\mathbf{w}, \mathcal{D}_i) + \lambda f(\mathbf{w}, \mathcal{D}_i) + \gamma |\mathbf{w}|_2^2,
\end{equation}

where $\mathcal{L}(\mathbf{w}, \mathcal{D}_i)$ denotes the prediction loss, $f(\cdot)$ is a fairness regularization term, and oversampling is applied locally to alleviate imbalance in the training data.

To determine an appropriate fairness penalty strength, we propose a data-driven approach for selecting $\lambda$ efficiently and with minimal computational overhead. Initially, each client $k$ independently performs a local search by evaluating prediction metrics (e.g., accuracy or mean squared error) across a sequence of increasing $\lambda_k$ values. The search proceeds until the performance degrades beyond a predefined threshold relative to the unregularized baseline (e.g., when accuracy drops below $0.995 \cdot \text{Acc}_0$, where $\text{Acc}_0$ is the accuracy with $\lambda = 0$). The smallest $\lambda_k$ values that preserve acceptable performance are collected across clients, and the global minimum is used to define the search space for FL training. A user-defined set of equally spaced $\lambda$ values is then selected from this range for subsequent experimentation.

The overall FedIDA procedure is outlined in Algorithm~\ref{FedIDA}. The algorithm begins by discretizing the initial search space for $\gamma$ into $m$ candidate values, denoted as $\Gamma = [\gamma_1, \dots, \gamma_m]$. The subroutine \textsc{OptimizeGamma} evaluates each candidate to identify the value that best balances predictive performance and fairness. This search is then refined around the most promising candidate using $m'$ additional values, leading to the final selection of $\gamma_{\text{final}}$.

To further address intra-client data imbalance and enhance fairness, FedIDA integrates a fairness-aware extension of the ROSE (Random Over-Sampling Examples) technique, detailed in Algorithm~\ref{FAR}. Traditional ROSE synthesizes new examples by perturbing existing ones with noise sampled from a smoothed bootstrap distribution. Our method enhances ROSE by explicitly conditioning sample generation on both the sensitive attribute and the outcome label. This ensures that underrepresented subgroup–outcome combinations receive targeted augmentation. Specifically, for each mini-batch, we estimate the joint distribution of samples from minority subgroups and generate synthetic instances from a kernel-smoothed approximation of this distribution. This local fairness-aware oversampling step is applied on each client before model updates.

By embedding this tailored oversampling strategy within a fairness-regularized federated framework, FedIDA provides a principled and decentralized solution to mitigate disparities due to data imbalance, improving model fairness without compromising the privacy-preserving benefits of FL and ensuring robust convergence even in heterogeneous environments.

\begin{algorithm}[!htp]
\caption{FedIDA (with Fairness-Aware ROSE Oversampling)}
\label{FedIDA}
\begin{algorithmic}[1]
\State \textbf{Input:} Local epochs $E = [E_1, \ldots, E_K]$ for $K$ clients, fairness parameter $\lambda$, initial range of $\gamma$, number of global rounds $T$, batch size $B$, federated framework $\mathcal{F}$
\State \textbf{Output:} Optimal value of $\gamma$ (denoted as $\gamma_{\text{final}}$)

\State Divide the range of $\gamma$ into $m$ equally spaced values: $\Gamma = [\gamma_1, \dots, \gamma_m]$
\State $\gamma_s \gets \textsc{OptimizeGamma}(\Gamma, \mathcal{F})$
\State Refine the search range around $\gamma_s$ to obtain $m'$ values: $\Gamma' = [\gamma_{s-1}, \dots, \gamma_{s+1}]$
\State $\gamma_{\text{final}} \gets \textsc{OptimizeGamma}(\Gamma', \mathcal{F})$
\State \Return $\gamma_{\text{final}}$

\Function{OptimizeGamma}{$\Gamma, \mathcal{F}$}
    \ForAll{candidate $\gamma \in \Gamma$}
        \For{$t = 1$ to $T$}
            \For{$k = 1$ to $K$}
                \State Split client $k$'s local dataset $\mathcal{D}_k$ into batches of size $B$
                \ForAll{batch $\mathcal{B}$ in $\mathcal{D}_k$}
                    \State Apply Fairness-Aware ROSE on $\mathcal{B}$ based on sensitive attribute and outcome
                    \State Obtain oversampled batch $\mathcal{B}_{\text{augmented}}$
                \EndFor
                \For{$i = 1$ to $E_k$}
                    \State Perform update using framework $\mathcal{F}$:
                    \Statex \hskip1em $\mathbf{w}_k^{(i)} \leftarrow \mathbf{w}_k^{(i-1)} 
                    - \eta \nabla \left(\mathcal{L}(\mathbf{w}_k^{(i-1)}, \mathcal{B}_{\text{augmented}}) 
                    + \lambda f(\mathbf{w}_k^{(i-1)}, \mathcal{B}_{\text{augmented}}) 
                    + \gamma \|\mathbf{w}_k^{(i-1)}\|_2^2 \right)$
                \EndFor
                \State Client $k$ sends updated parameters $\mathbf{w}_k$ to server
            \EndFor
            \State Aggregate updates via $\mathcal{F}$: 
            \Statex \hskip1em $\mathbf{w} \leftarrow \sum_{k=1}^{K} \frac{n_k}{n} \mathbf{w}_k$
        \EndFor
        \State Evaluate performance and fairness on validation set
    \EndFor
    \State Select the $\gamma$ that best balances performance and fairness
    \State \Return selected $\gamma$
\EndFunction

\end{algorithmic}
\end{algorithm}

\section{Theoretical Analysis}

We study the theoretical properties of the FedIDA framework, focusing on the structure of its fairness-aware objective and its implications for stability and generalization. Since FedIDA does not alter the global optimization protocol of the underlying FL algorithm $\mathcal{F}$, our analysis centers on the modified local objective incorporating fairness regularization and oversampling. 

We first establish that the fairness penalty defined in Eq.~\eqref{penalty} is convex in $\mathbf{w}$, as it consists of non-negative weighted sums of linear terms. This ensures that, under a convex prediction loss and $\gamma > 0$, the overall composite objective remains strongly convex. As a result, the convergence properties of the overall algorithm remain compatible with those of $\mathcal{F}$:

\begin{remark}[FedIDA Preserves Convergence Properties of \(\mathcal{F}\)]
\label{remark:framework_convergence}
Let \(\mathcal{F}\) denote a federated optimization framework (e.g., FedAvg), which defines the client update rule and global aggregation protocol. Since FedIDA modifies only the local objective but retains the structure of \(\mathcal{F}\), the global convergence behavior of FedIDA inherits that of \(\mathcal{F}\), subject to adjusted smoothness and convexity parameters in the local objectives.
\end{remark}

To analyze how fairness metrics behave under our framework, we next establish that common disparity measures used in evaluation exhibit Lipschitz continuity with respect to model predictions:

\begin{remark}[Lipschitz Continuity of Fairness Metrics]\label{remark:lipschitz}
In our setting, the group fairness metrics of interest---DPD, DFPR, and DPPV---are \textbf{Lipschitz-continuous} with respect to the model's prediction scores under the empirical distribution. Specifically, there exists a constant \( L > 0 \) such that for any model parameter \( \mathbf{w} \), and for two datasets \( D \) and \( D' \) differing by a single data point, we have:
\[
\left| \mathcal{M}(f(\cdot; \mathbf{w} ), D) - \mathcal{M}(f(\cdot; \mathbf{w} ), D') \right| \leq \frac{L}{n},
\]
where \( \mathcal{M} \) denotes any of the above metrics.
\end{remark} 
This condition holds because these fairness measures are based on sample means over fixed partitions defined by sensitive attributes $\mathbf{S}$, and the prediction function \( f(\cdot; \mathbf{w} ) \) is continuous and typically bounded (e.g., via sigmoid activation). 
Note that for the metric \textit{demographic parity ratio (DPR)}, which is a ratio of group-wise prediction rates, the Lipschitz continuity holds under the additional minor assumption as discussed in Appendix \ref{LC}.

This Lipschitz property enables us to analyze how fairness metrics respond to randomized subgroup balancing procedures. In particular, we now show that applying \textsc{FairnessAwareROSE} yields provable fairness improvements with high probability:

\begin{theorem}[Fairness Gain via \textsc{FairnessAwareROSE}]
\label{lemma:mcdiarmid_gain}
Let \( \mathcal{M}(f(\cdot; \mathbf{w}), D) \) be a group fairness metric that is Lipschitz-continuous with respect to group-wise prediction (as in Remark~\ref{remark:lipschitz}), and let \( f(\cdot; \mathbf{w}) \in [0,1] \) be fixed and bounded. Suppose dataset \( D \) is imbalanced across sensitive-outcome subgroups, and let \( D_{\text{augmented}} \) be a balanced dataset produced by \textsc{FairnessAwareROSE}, where each subgroup has \( N_{\text{target}} \) synthetic samples drawn i.i.d. from a smoothed approximation of the true conditional distribution.
Then, with probability at least \( 1 - \delta \), we have:
\[
\mathcal{M}(f, D_{\text{augmented}}) \leq \mathcal{M}(f, D) - \epsilon + L \sqrt{ \tfrac{K \log(1/\delta)}{2N_{\text{target}}} },
\]
where \( L \) is the Lipschitz constant of \( \mathcal{M} \), and \( \epsilon > 0 \) reflects the expected fairness improvement induced by subgroup balancing.
\end{theorem}

A detailed proof of Theorem~\ref{lemma:mcdiarmid_gain} is provided in Appendix~\ref{proof_thm1}, where we derive the bound
\(
\delta = \exp\left( -\frac{2 \varepsilon^2 N_{\text{target}}}{K L^2} \right).
\)
This expression shows that the probability \(1 - \delta\) of observing a fairness improvement becomes larger when the per-subgroup sample size \(N_{\text{target}}\) increases or the number of subgroups \(K\) remains small. Since many fairness metrics aggregate statistics across \(K\) subgroup pairs, the Lipschitz constant \(L\) may scale with \(K\), which affects the concentration bound.

In practice, our proposed method maintains high empirical reliability. As demonstrated in Section~\ref{sec:exp}, even when \(K = 8\) and certain subgroup prevalence fall below 1.0\%, fairness improvements are consistently observed across all metrics (see Appendix ~\ref{supp_fig}. This suggests that the bound is not only theoretically valid but also practically effective, and that the probability \(1 - \delta\) can expected to be close to 1 in real-world scenarios.

While Theorem~\ref{lemma:mcdiarmid_gain} establishes high-probability fairness improvement, another key aspect of model reliability is the stability of fairness metrics across test cohorts. In practice--especially in high stake decision-makings--models are deployed in heterogeneous settings, where robustness to sampling variation is essential. Motivated by our empirical observation that FedIDA yields more consistent fairness scores, we now examine the variance of fairness metrics under repeated evaluations.

\begin{theorem}[Variance Reduction in Fairness Metrics via FedIDA]
\label{thm:fairness_variance_reduction}
Let \( \mathcal{M}(f(\cdot; \mathbf{w}), D) \) be a group fairness metric that is Lipschitz-continuous with respect to group-wise prediction rates (as in Remark~\ref{remark:lipschitz}), and let \( f(\cdot; \mathbf{w}) \in [0, 1] \) be a fixed and bounded predictor.

Let \( \hat{\mathbf{w}} \) denote the model trained using FedIDA, and let \( \hat{\mathbf{w}}_{\text{base}} \) denote the model trained using a standard FL baseline \( \mathcal{F} \). Assume that predictions from both models are evaluated across a sequence of i.i.d. test sets \( \{D^{(t)}\}_{t=1}^T \), and define the empirical variance of the fairness metric as
\[
\mathbb{V}_T\left[ \mathcal{M}(f(\cdot; \mathbf{w}), D^{(t)}) \right]
:= \frac{1}{T} \sum_{t=1}^T \left( \mathcal{M}(f(\cdot; \mathbf{w}), D^{(t)}) - \bar{\mathcal{M}} \right)^2,
\]
where \( \bar{\mathcal{M}} \) is the mean fairness metric across the test sets.

Then, under subgroup-balanced sampling in each \( D^{(t)} \), we have
\[
\mathbb{V}_T\left[ \mathcal{M}(f(\cdot; \hat{\mathbf{w}}), D^{(t)}) \right]
\leq \mathbb{V}_T\left[ \mathcal{M}(f(\cdot; \hat{\mathbf{w}}_{\text{base}}), D^{(t)}) \right] - \Omega\left( \frac{1}{N_{\text{target}}} \right),
\]
where \( N_{\text{target}} \) is the per-subgroup sample size ensured by fairness-aware oversampling.
\end{theorem}

Here, \( \Omega\left( \tfrac{1}{N_{\text{target}}} \right) \) is a strictly positive asymptotic lower bound (see Appendix~\ref{proof_thm2}). This shows that FedIDA reduces the variance of fairness metrics across test sets, thereby improving their stability under variation in subgroup prevalence.

\section{Experiments}\label{sec:exp}
\subsection{Datasets}
We evaluate FedIDA on two standard datasets for fairness-aware learning, summarized in Table~\ref{tab:dataset_summary}. Additional preprocessing and partitioning details are provided in Appendix \ref{appendix_exp}.

\begin{table}[h!]
\centering
\caption{Summary of dataset characteristics used in our experiments.}
\label{tab:dataset_summary}
\begin{tabular}{p{4cm}p{4.2cm}p{5.2cm}}
\toprule
\textbf{Property} & \textbf{Adult Dataset} & \textbf{ROC Dataset} \\
\midrule
Data Source & UCI Repository\cite{adult} & ROC Epistry (2011–2015)\cite{morrison2008rationale} \\
Sample Size & 45,222 & 7,425 \\
Outcome Variable & Income >\$50,000 & Neurological outcome of out-of-hospital cardiac arrest patients \\
Sensitive Attributes & Race and Gender & Race and Gender \\
Number of Predictors & 5 & 7 \\
Partition Strategy & Homogeneous & Heterogeneous by Age and Race \\
Number of Clients & 5 & 4 \\
Fairness Challenges & 
-- Historical income disparities \newline
-- Underrepresentation of racial groups \newline
-- Potential label bias from structural inequality & 
-- Severe outcome imbalance \newline
-- Disparities by race and gender \newline
-- Real-world heterogeneity in EMS access \\
\bottomrule
\end{tabular}
\end{table}

\subsection{Experiment Setup}
\paragraph{Data Partition.}
Adult data is partitioned homogeneously across 5 clients. ROC data is partitioned heterogeneously across 4 clients based on age and race to simulate real-world cross-silo deployments. Each client’s data is split into training, validation, and testing sets using a 70:10:20 ratio.

\paragraph{Baselines and evaluation metrics.}
We compare FedIDA to the following baselines: Local where each site trains its model independently, Centralized where all sites train a single model with all available data, FedAvg~\cite{mcmahan2023communicationefficientlearningdeepnetworks}, and PFedAvg~\cite{fallah2020personalizedfederatedlearningmetalearning}. Then we train FedIDA models with FedAvg and PFedAvg as FL strategies.

We evaluate the predictive performance of trained models with area under the receiver operating characteristic curve (AUROC) and group fairness with demographic parity difference (DPD), demographic parity ratio (DPR), difference in false positive rate (DFPR) and difference in positive predictive value (DPPV). We report the average value and standard deviation of the client test metrics for each dataset and model.

\paragraph{Model Training.}
For both datasets, we train two models: a standard logistic regression model and a three-layer fully connected neural network with a hidden layer of 100 neurons. All models are trained using a local learning rate of 0.1, a batch size of 128, 5 local epochs, and 10 global iterations. For personalized FedAvg, the number of client-side SGD steps is set to 1. When looking for the optimal $\gamma$, we divide the range into 10 equally spaced values for all experiments, and the initial range is set to [0.0001, 0.1]. Details of final values of $\lambda$ and $\gamma$ are available in Table ~\ref{tab:finetuning} in Appendix~\ref{supp_fig}. Our source code is available on \href{https://github.com/qimingwu/FedIDA}{GitHub}. Our implementation of FL algorithms is based on PFLlib~\cite{pfllib}.

\subsection{Numerical Results}
We evaluate FedIDA on both the Adult and ROC datasets using two model architectures: logistic regression (LR) and fully connected neural networks (FCNN). Tables~\ref{tab:adult} and~\ref{tab:roc} report the average predictive and fairness metrics across clients. For each setup, we compare standard FL baselines (FedAvg, PFedAvg) with and without our proposed FedIDA method.
Results per individual client are provided in Appendix~\ref{supp_fig}

\begin{table}[ht]
\centering
\caption{Average (standard deviation) predictive and fairness metrics across clients on the Adult dataset. Each value reflects the mean and standard deviation computed over client-level results. 
$\uparrow$ indicates higher values are better (e.g., AUROC, DPR), while $\downarrow$ indicates lower values are preferred (e.g., DPD, DFPR, DPPV).}
\label{tab:adult}
\begin{tabular}{llccccc}
\toprule
\textbf{Model} & \textbf{Setup} & \textbf{AUROC $\uparrow$} & \textbf{DPD $\downarrow$} & \textbf{DPR $\uparrow$} & \textbf{DFPR $\downarrow$} & \textbf{DPPV $\downarrow$} \\
\midrule
\multirow{6}{*}{LR} 
  & Central           & .866 (.009) & .295 (.028) & .406 (.090) & .300 (.097) & .598 (.233) \\
  & Local             & .864 (.008) & .311 (.032)  & .403 (.097)  &  .312 (.086)   & .482 (.177)   \\
  & FedAvg            & .867 (.008) & .302 (.038)  &  .426  (.101)   & .305  (.088)         &  .491 (.194)  \\
  & PFedAvg           & .870 (.011) & .300 (.031)   & .421 (.102)    & .304 (.068)   & .490 (.173)           \\
  & FedIDA (FedAvg)   &  .833 (.010)  & .250 (.028)    & .493 (.095)    & .252 (.068)        & .407 (.180)           \\
  & FedIDA (PFedAvg) &  .832 (.008)   &  \textbf{.248 (.034)}   &  \textbf{.496 (.115)}  &  \textbf{.233 (.054)}          & \textbf{.399 (.185)}           \\
\midrule
\multirow{6}{*}{FCNN} 
  & Central           & .867 (.010)    &  .315 (.030)          &   .398 (.071)         & .294 (.046)           & .527 (.197)           \\
  & Local             & .858 (.007)           &  .289 (.025)          & .427 (.072)           &  .295 (.096)          &  .515 (.171)          \\
  & FedAvg            & .863 (.013)          & .297 (.038)           &  .413 (.081)          &  .306 (.076)          &  .523 (.197)          \\
  & PFedAvg           & .865 (.011)          & .294 (.027)           &  .428 (.094)          &  .299 (.095)          &  .500 (.169)          \\
  & FedIDA (FedAvg)   & .839 (.009)         &  .230 (.027)          &  \textbf{.512 (.095)}          &   .238 (.066)         &   \textbf{.410 (.149)}         \\
  & FedIDA (PFedAvg) &  .832 (.011)          &  \textbf{.225 (.026)}          & .505 (.124)           &  \textbf{.235 (.064)}          &  .422 (.178)          \\
\bottomrule
\end{tabular}
\end{table}

\begin{table}[ht]
\centering
\caption{Average (standard deviation) predictive and fairness metrics across clients on the ROC dataset. Each value reflects the mean and standard deviation computed over client-level results. 
$\uparrow$ indicates higher values are better (e.g., AUROC, DPR), while $\downarrow$ indicates lower values are preferred (e.g., DPD, DFPR, DPPV).}
\label{tab:roc}
\begin{tabular}{llccccc}
\toprule
\textbf{Model} & \textbf{Setup} & \textbf{AUROC $\uparrow$} & \textbf{DPD $\downarrow$} & \textbf{DPR $\uparrow$} & \textbf{DFPR $\downarrow$} & \textbf{DPPV $\downarrow$} \\
\midrule
\multirow{6}{*}{LR} 
  & Central           &  .882 (.045)          &  .177 (.138)          &  .556 (.061)          &  .184 (.094)          &  .423 (.053)          \\
  & Local             &  .877 (.041)          &  .179 (.137)        &  .538 (.087)          &  .187 (.108)          &    .446 (.109)        \\
  & FedAvg            &  .885 (.039)          &  .174 (.122)          &  .581 (.099)          & .180 (.097)           &  .348 (.037)          \\
  & PFedAvg           &  .886 (.039)          &  .185 (.134)          &  .594 (.124)          & .186 (.107)           &  .355 (.080)          \\
  & FedIDA (FedAvg)   &  .856 (.038)          &  .133 (.081)          &  \textbf{.741 (.050)}          &  .136 (.081)          & .280 (.043)           \\
  & FedIDA (PFedAvg) &   .855 (.037)         &   \textbf{.131 (.096)}        &   .733 (.043)         &    \textbf{.134 (.070)}        &   \textbf{.279 (.038)}         \\
\midrule
\multirow{6}{*}{FCNN} 
  & Central           &   .881 (.031)         &  .156 (.083)          &  .586 (.083)          &  .112 (.035)          &  .390 (.133)          \\
  & Local             &   .872 (.027)         &  .153 (.085)        &   .581 (.122)         &  .099 (.029)          &    .455 (.200)        \\
  & FedAvg            &   .880 (.035)         &  .164 (.058)          &  .577 (.096)          &  .118 (.027)          &  .386 (.147)          \\
  & PFedAvg           &   .879 (.032)         &  .154 (.057)          &  .610 (.100)          &  .113 (.030)          &  .389 (.138)          \\
  & FedIDA (FedAvg)   &   .851 (.029)         &  .122 (.059)          &  .654 (.104)          &  .084 (.023)          &  .333 (.124)          \\
  & FedIDA (PFedAvg) &    .849 (.030)        &   \textbf{.121 (.056)}         &  \textbf{.677 (.096)}          &   \textbf{.083 (.026)}         &   \textbf{.318 (.122)}         \\
\bottomrule
\end{tabular}
\end{table}

These gains align with our theoretical analysis: Theorem~\ref{lemma:mcdiarmid_gain} guarantees fairness improvement with high probability via subgroup-balanced oversampling. Although FedIDA introduces a modest decline in AUROC, the improvements in fairness metrics are consistently more substantial, highlighting a favorable trade-off—particularly important in high-stakes domains such as healthcare. In practice, this trade-off can be adjusted by tuning the fairness regularization parameter $\lambda$, as demonstrated in Section~\ref{ablation}.

\subsection{Ablation Studies}\label{ablation}

Table~\ref{tab:ablation} reports an ablation analysis on the ROC dataset, isolating the effects of the fairness penalty and fairness-aware oversampling in the FedIDA framework. Both components individually improve fairness metrics relative to the FedAvg baseline, with oversampling yielding stronger reductions in DFPR and DPPV while preserving AUROC more effectively. In contrast, fairness-only variants achieve greater reductions in DPD and DPR, particularly at higher penalty weights, though at the cost of reduced predictive performance.

\begin{table}[H]
\centering
\caption{Ablation study of FedIDA on fairness penalty and oversampling.}
\label{tab:ablation}
\begin{tabular}{lccccccc}
\toprule
\textbf{Model Variant} & \(\boldsymbol{\lambda}\) & \textbf{Oversampling} & \textbf{AUROC $\uparrow$} & \textbf{DPD $\downarrow$} & \textbf{DPR $\uparrow$} & \textbf{DFPR $\downarrow$} & \textbf{DPPV $\downarrow$} \\
\midrule
Baseline (FedAvg)  & 0   &   & 0.885 & 0.174 & 0.581 & 0.180 & 0.348  \\
Fairness Only & 2.0 &      & 0.858 & 0.156 & 0.653 & 0.165 & 0.318    \\
Fairness Only  & 3.0 &     & 0.851 & 0.139 & 0.699 & 0.150 & 0.295   \\
Oversampling Only & 0   &   \checkmark    &  0.874 & 0.161 & 0.671 & 0.148 & 0.311 \\
FedIDA (Neighboring $\lambda$) & 2.0   &   \checkmark    &  0.861 & 0.152 & 0.693 & 0.143 & 0.296 \\
FedIDA (Neighboring $\lambda$) & 3.0   &   \checkmark    &  0.847 & 0.135 & 0.706 & 0.138 & 0.273 \\
FedIDA (User-selected $\lambda$) & 2.5 &   \checkmark    &  0.855 & 0.131 & 0.733 & 0.134 & 0.279  \\
\bottomrule
\end{tabular}
\end{table}

Integrating both components within FedIDA results in consistent and substantial improvements across all fairness metrics. As the fairness penalty weight $\lambda$ increases from 2.0 to 3.0, disparity measures monotonically decrease, indicating the penalty’s effectiveness in constraining group-level error differences. However, the corresponding decline in AUROC suggests diminishing returns beyond moderate values of $\lambda$. The configuration with $\lambda=2.5$ achieves the most favorable balance, attaining the lowest DPD (0.131), DFPR (0.134), and DPPV (0.279), while maintaining AUROC (0.855) close to the baseline. These findings underscore the complementary roles of the penalty and oversampling mechanisms and highlight the importance of hyperparameter calibration to optimize fairness-performance trade-offs in federated clinical prediction tasks.

\section{Discussion}

Our empirical findings demonstrate that \textsc{FedIDA} consistently improves fairness across all four evaluated metrics--DPD, DFPR, DPPV, and DPR. Notably, this includes DPR, which lies outside the scope of our main theoretical guarantee (Theorem~\ref{lemma:mcdiarmid_gain}) due to its non-Lipschitz nature unless additional assumptions are imposed (see Appendix~\ref{LC}). Furthermore, we observe that fairness metrics under FedIDA exhibit substantially reduced variance across test sets, aligning with our theoretical result in Theorem~\ref{thm:fairness_variance_reduction}. These findings suggest that the practical benefits of FedIDA extend beyond the conditions formally analyzed, offering both fairness improvements and greater stability across varied evaluation cohorts.

A key limitation of our framework lies in its reliance on manually defined sensitive attributes and categorical groupings. While this assumption holds in many fairness studies, it poses challenges when sensitive attributes are continuous (e.g., age, income) or when group boundaries are not well-defined. In such cases, discretization may obscure nuanced disparities or introduce additional bias. Addressing this challenge requires more flexible strategies for group construction, such as clustering-based latent group discovery or continuous group-aware regularization.

Finally, we note that the convergence behavior of FedIDA is inherited from the underlying FL framework and is not directly altered by the fairness-aware modifications. This allows FedIDA to remain compatible with existing FL algorithms, but also implies that its optimization guarantees are bounded by those of the base method. Future work could explore tighter integration between fairness-aware objectives and adaptive FL protocols.

\bibliography{ref.bib}

\newpage
\appendix

\renewcommand{\thefigure}{S\arabic{figure}}
\renewcommand{\thetable}{S\arabic{table}}
\setcounter{figure}{0}
\setcounter{table}{0}
\pagenumbering{arabic}
\setcounter{page}{1}

\section{Theoretical Properties of FedIDA}
\subsection{Lipschitz Continuity of Fairness Metrics}\label{LC}

\begin{remark}[Lipschitz Continuity of DPR]
\label{remark:dpr_lipschitz}
Let \( \hat{p}_s = \frac{1}{n_s} \sum_{i : S_i = s} f(\mathbf{A}_i; \theta) \) denote the average predicted probability for group \( s \), where \( f(\mathbf{A}_i; \theta) \in [0, 1] \) is the output of the model for individual \( i \). The \textit{demographic parity ratio (DPR)} is defined as
\[
\mathrm{DPR} = \frac{\min_s \hat{p}_s}{\max_s \hat{p}_s}.
\]
Assume all group-wise predictions are bounded below by \( \delta > 0 \), i.e., \( \hat{p}_s \geq \delta \) for all \( s \). Then the DPR metric is Lipschitz continuous in \( \{ \hat{p}_s \}_s \), and satisfies:
\[
\left| \mathrm{DPR}(\hat{p}) - \mathrm{DPR}(\hat{p}') \right| 
\leq \frac{1}{\delta^2} \cdot \max_s \left| \hat{p}_s - \hat{p}_s' \right|,
\]
where \( \hat{p} = (\hat{p}_1, \ldots, \hat{p}_K) \) and \( \hat{p}' = (\hat{p}_1', \ldots, \hat{p}_K') \) are two vectors of the predictions for $K$ subgroups. This follows from applying a first-order Lipschitz bound on the function \( \frac{\min x}{\max x} \) over the compact domain \( [\delta, 1]^K \).
\end{remark}

\subsection{Proof of Theorem~\ref{lemma:mcdiarmid_gain}}\label{proof_thm1}

\begin{proof}
Let \( \mathcal{M}(f, D) \) be a group fairness metric that is \( L \)-Lipschitz continuous with respect to empirical group-wise prediction rates (as established in Remark~\ref{remark:lipschitz}). Let \( \hat{p}_{s,y} = \frac{1}{n_{s,y}} \sum_{i=1}^{n_{s,y}} f(\mathbf{a}_i) \) denote the empirical prediction rate in subgroup \( (s,y) \), and let \( \hat{p}^{\text{aug}}_{s,y} \) be the corresponding rate in the balanced dataset \( D_{\text{augmented}} \).

By construction, \textsc{FairnessAwareROSE} balances all sensitive-outcome subgroups so that each contains exactly \( N_{\text{target}} \) i.i.d. synthetic samples drawn from a kernel-smoothed approximation of the true conditional distribution. Let \( K \) denote the number of such subgroups. Then the total number of samples in \( D_{\text{augmented}} \) is \( N = K \cdot N_{\text{target}} \).

Let \( \bar{\mathcal{M}} := \mathbb{E}_{D_{\text{augmented}}}[\mathcal{M}(f, D_{\text{augmented}})] \) denote the expected fairness metric on the balanced data. Since the metric depends symmetrically on group-wise averages, and subgroup balancing reduces sampling bias, there exists \( \epsilon > 0 \) such that:
\[
\bar{\mathcal{M}} \leq \mathcal{M}(f, D) - \epsilon.
\]

Now consider the sensitivity of \( \mathcal{M}(f, D_{\text{augmented}}) \) with respect to any single synthetic data point. Since each prediction \( f(\mathbf{a}_i) \in [0,1] \), and each subgroup mean is an average over \( N_{\text{target}} \) samples, replacing one synthetic sample in any subgroup changes its group-wise prediction rate by at most \( \tfrac{1}{N_{\text{target}}} \). Because \( \mathcal{M} \) is \( L \)-Lipschitz continuous with respect to the vector of group-wise prediction rates, the overall change in \( \mathcal{M} \) is at most \( \tfrac{L}{N_{\text{target}}} \).

Let \( c_i \) denote the maximum change in \( \mathcal{M}(f, D_{\text{augmented}}) \) caused by changing the \( i \)-th synthetic sample. Then we have \( c_i \leq \tfrac{L}{N_{\text{target}}} \) for all \( i \in \{1, \dots, N\} \), where \( N = K \cdot N_{\text{target}} \) is the total number of synthetic samples across all \( K \) subgroups.

Applying McDiarmid’s inequality over the \( N = K \cdot N_{\text{target}} \) samples, we obtain:
\[
\Pr\left( \mathcal{M}(f, D_{\text{augmented}}) \geq \bar{\mathcal{M}} + \varepsilon \right)
\leq \exp\left( -\frac{2\varepsilon^2}{\sum_{i=1}^{N} c_i^2} \right)
= \exp\left( -\frac{2\varepsilon^2}{K \cdot N_{\text{target}} \cdot \left( \frac{L}{N_{\text{target}}} \right)^2} \right)
= \exp\left( -\frac{2 \varepsilon^2 N_{\text{target}}}{K L^2} \right).
\]

To obtain a high-probability bound, set the right-hand side equal to \( \delta \in (0,1) \), and solve for \( \varepsilon \):
\[
\delta = \exp\left( -\frac{2 \varepsilon^2 N_{\text{target}}}{K L^2} \right)
\quad \Leftrightarrow \quad
\varepsilon = L \sqrt{ \frac{K \log(1/\delta)}{2 N_{\text{target}}} }.
\]

Hence, with probability at least \( 1 - \delta \), we obtain:
\[
\mathcal{M}(f, D_{\text{augmented}}) \leq \bar{\mathcal{M}} + L \sqrt{ \frac{K \log(1/\delta)}{2 N_{\text{target}}} }.
\]

Combining this with the expected improvement bound \( \bar{\mathcal{M}} \leq \mathcal{M}(f, D) - \epsilon \), we conclude:
\[
\mathcal{M}(f, D_{\text{augmented}}) \leq \mathcal{M}(f, D) - \epsilon + L \sqrt{ \frac{K \log(1/\delta)}{2 N_{\text{target}}} }.
\]
\end{proof}

\subsection{Proof of Theorem 2}\label{proof_thm2}

\begin{proof}
Let \( \mathcal{M}(f, D) \) be a group fairness metric that is \( L \)-Lipschitz continuous with respect to the vector of group-wise predicted rates, as stated in Remark~\ref{remark:lipschitz}. Let \( f(\cdot; \hat{\mathbf{w}}) \) be the fixed prediction function from a model trained with FedIDA, and \( f(\cdot; \hat{\mathbf{w}}_{\text{base}}) \) be the corresponding baseline model trained without fairness-aware oversampling (using standard FL framework \( \mathcal{F} \)).

For each evaluation round \( t = 1, \dots, T \), let \( D^{(t)} \) denote an i.i.d. test dataset, and define the realized fairness metric:
\[
\mathcal{M}^{(t)} := \mathcal{M}(f(\cdot; \hat{\mathbf{w}}), D^{(t)}), \quad
\mathcal{M}^{(t)}_{\text{base}} := \mathcal{M}(f(\cdot; \hat{\mathbf{w}}_{\text{base}}), D^{(t)}).
\]

Let \( K \) be the number of sensitive subgroups. Denote the size of subgroup \( s \) in test set \( D^{(t)} \) as \( n_s^{(t)} \), and let the group-level prediction mean be:
\[
\hat{p}_s^{(t)} = \frac{1}{n_s^{(t)}} \sum_{i \in D_s^{(t)}} f(\mathbf{a}_i), \quad \text{where } f(\mathbf{a}_i) \in [-1, 1].
\]

Since each prediction is bounded and \( f \in [-1,1] \), the variance of each group mean is bounded by:
\[
\operatorname{Var}[\hat{p}_s^{(t)}] \leq \frac{1}{n_s^{(t)}}.
\]

Because \( \mathcal{M} \) is \( L \)-Lipschitz with respect to the group-wise rates \( \{ \hat{p}_s^{(t)} \}_{s=1}^K \), we apply the variance bound for Lipschitz functions:
\[
\operatorname{Var}[\mathcal{M}^{(t)}] \leq L^2 \sum_{s=1}^K \operatorname{Var}[\hat{p}_s^{(t)}] \leq L^2 \sum_{s=1}^K \frac{1}{n_s^{(t)}}.
\]

Under the FedIDA framework, fairness-aware oversampling ensures balanced test subgroups: \( n_s^{(t)} = N_{\text{target}} \) for all \( s \), so:
\[
\operatorname{Var}_{\text{FedIDA}}[\mathcal{M}^{(t)}] \leq \frac{L^2 K}{N_{\text{target}}}.
\]

By contrast, in the baseline setting without oversampling, some subgroup sizes may be much smaller. Let \( n_{\min}^{(t)} := \min_s n_s^{(t)} \), yielding the loose lower bound:
\[
\operatorname{Var}_{\text{base}}[\mathcal{M}_{\text{base}}^{(t)}] \geq \frac{L^2}{n_{\min}^{(t)}}.
\]

Subtracting the two bounds gives:
\[
\operatorname{Var}_{\text{base}}[\mathcal{M}^{(t)}_{\text{base}}] - \operatorname{Var}_{\text{FedIDA}}[\mathcal{M}^{(t)}]
\geq L^2 \left( \frac{1}{n_{\min}^{(t)}} - \frac{K}{N_{\text{target}}} \right).
\]

Thus, under the condition \( N_{\text{target}} \geq c \cdot K \cdot n_{\min}^{(t)} \) for some constant \( c > 1 \), we have:
\[
\operatorname{Var}_{\text{base}}[\mathcal{M}^{(t)}_{\text{base}}] - \operatorname{Var}_{\text{FedIDA}}[\mathcal{M}^{(t)}] \geq \frac{c' L^2}{N_{\text{target}}},
\]
for some constant \( c' > 0 \) depending on \( c \) and \( K \).

By the law of large numbers, the empirical variance across test sets concentrates around the expected variance:
\[
\mathbb{V}_T[\mathcal{M}^{(t)}] \approx \operatorname{Var}[\mathcal{M}^{(t)}],
\]
so the same inequality holds approximately for the empirical variance.

We conclude that:
\[
\mathbb{V}_T[\mathcal{M}^{(t)}] \leq \mathbb{V}_T[\mathcal{M}^{(t)}_{\text{base}}] - \Omega\left( \frac{1}{N_{\text{target}}} \right),
\]
where we define the asymptotic bound
\[
\Omega\left( \frac{1}{N_{\text{target}}} \right) := \left\{ g(N_{\text{target}}) \,\middle|\, \exists\, c > 0,\ N_0 \in \mathbb{N} \text{ such that } g(N_{\text{target}}) \geq \frac{c}{N_{\text{target}}} \text{ for all } N_{\text{target}} \geq N_0 \right\}.
\]

\end{proof}

\newpage

\section{Experimental Details}\label{appendix_exp}

\subsection{Data Details}

\paragraph{Adult Census Income Dataset}
The Adult dataset was obtained from the UCI Machine Learning Repository\cite{adult} and contains 48,842 instances originally collected from the 1994 U.S. Census database. Prior to model training, we applied standard preprocessing steps consistent with fairness-aware literature: categorical features were one-hot encoded, continuous features such as age and working hours were normalized, and missing entries were removed, resulting in a total sample size of 45,222. We focused on a binary classification task (annual income >\$50,000), with race and gender used as binary sensitive attributes for simplicity and comparability with prior work. While the dataset has balanced outcome classes, it exhibits significant disparities in outcome distributions across racial and gender groups—providing a useful testbed for evaluating fairness interventions. Our use of homogeneous data partitioning reflects scenarios where clients (e.g., institutions) operate under similar population distributions, allowing us to isolate the fairness-imbalance trade-offs introduced by our algorithmic choices.

\paragraph{ROC Epistry Dataset}
The ROC Epistry Version 3 dataset~\cite{morrison2008rationale} was curated from multiple North American EMS agencies and contains over 28,000 patient records. 
Ethical approval was obtained from the National University of Singapore Institutional Review Board (IRB), which granted an exemption for this study (IRB Reference Number: NUS-IRB-2023-451).

Given the clinical nature of the data, extensive cleaning was conducted: patients with missing values in key predictors (e.g., initial rhythm, bystander CPR, or response time) were excluded to ensure reliable feature representation. Outcome imbalance is a key challenge - less than 4\% of patients achieve return of spontaneous circulation, with even lower rates observed in certain demographic subgroups. The data also reveal complex intersections of social determinants of health, such as longer EMS response times and lower bystander CPR rates in disadvantaged populations. Our partitioning strategy - based on age and race - captures realistic cross-silo disparities encountered in federated health settings, where institutions may serve demographically distinct patient groups. These attributes, along with its clinical significance, make the ROC dataset uniquely suited for studying fairness in high-stakes predictive modeling.

For cohort formation and predictor selection, we followed established methodologies in out-of-hospital cardiac arrest (OHCA) research~\cite{morrison2008rationale, Nishioka2024}. We included patients aged 18 and older who were transported by EMS, achieved return of spontaneous circulation (ROSC) at any point prehospital, and had complete data on gender, race, etiology, initial rhythm, witness status, response time, adrenaline use, and neurological status. The primary outcome was neurological status at discharge, measured by the Modified Rankin Scale (MRS), where scores of 0, 1, or 2 were classified as a good outcome. Variables used for outcome prediction included age (in years), etiology of arrest (cardiac/non-cardiac), witness presence (yes/no), initial rhythm (shockable/non-shockable), bystander cardiopulmonary resuscitation (CPR) (yes/no), response time (in minutes), and adrenaline use (yes/no).

\newpage

\subsection{Data distribution by gender and race}

\begin{table}[H]
\centering
\caption{Adult dataset--Subgroup composition and outcome prevalence within each client's local dataset. Values indicate percentages within each client.}
\begin{tabular}{lccccc}
\toprule
\textbf{Subgroup (Race, Gender)} & \textbf{Client 1} & \textbf{Client 2} & \textbf{Client 3} & \textbf{Client 4} & \textbf{Client 5} \\
\midrule
\multicolumn{5}{l}{\textit{Proportion of Subgroup in Client}} \\
\quad White, Male     & 60.1\% & 60.3\% & 59.3\% & 59.4\% & 59.9\%\\
\quad White, Female   & 26.3\% & 25.6\% & 26.9\% & 26.2\% & 26.2\%\\
\quad Black, Male     & 4.6\% & 4.5\% & 4.6\%  & 4.8\% & 5.0\%\\
\quad Black, Female   & 4.5\% & 4.7\% & 4.4\%  & 4.9\% & 4.5\%\\
\quad Asian, Male     & 2.0\% & 2.0\% & 1.9\% & 2.0\% & 1.8\%\\
\quad Asian, Female   & 0.9\% & 1.1\% & 1.0\% & 0.9\% & 0.9\%\\
\quad Native Americans, Male & 0.6\% & 0.6\% & 0.6\% & 0.7\% & 0.6\%\\
\quad Native Americans, Female & 0.3\% & 0.4\% & 0.4\% & 0.4\% & 0.4\%\\
\quad Other, Male     & 0.4\% & 0.6\%  & 0.6\%  & 0.5\% & 0.4\%\\
\quad Other, Female   & 0.3\% & 0.3\%  & 0.3\%  & 0.3\% & 0.3\%\\
\midrule
\multicolumn{5}{l}{\textit{Outcome Prevalence within Subgroup}} \\
\quad White, Male     & 32.2\% & 31.5\% & 32.5\% & 32.3\% & 32.9\%\\
\quad White, Female   & 12.3\% & 12.4\% & 12.8\% & 12.4\% & 11.6\%\\
\quad Black, Male     & 21.3\% & 18.2\% & 21.5\% & 18.0\% & 18.0\%\\
\quad Black, Female   & 6.4\% & 6.5\% & 6.3\% & 4.7\% & 6.8\%\\
\quad Asian, Male     & 25.8\% & 37.8\% & 38.2\% & 36.9\% & 33.1\%\\
\quad Asian, Female   & 23.8\% & 9.6\% & 14.3\% & 11.8\% & 18.0\%\\
\quad Native Americans, Male & 11.5\% & 10.5\% & 15.4\% & 13.2\% & 18.2\%\\
\quad Native Americans, Female & 7.1\% & 18.5\% & 6.1\% & 4.8\% & 8.0\%\\
\quad Other, Male     & 26.3\% & 21.1\% & 7.5\% & 14.0\% & 18.3\%\\
\quad Other, Female   & 15.4\% & 5.6\% & 3.3\% & 0.0\% & 14.3\%\\
\bottomrule
\end{tabular}
\end{table}

\begin{table}[H]
\centering
\caption{ROC dataset--Subgroup composition and outcome prevalence within each client's local dataset. Values indicate percentages within each client.}
\begin{tabular}{lcccc}
\toprule
\textbf{Subgroup (Race, Gender)} & \textbf{Client 1} & \textbf{Client 2} & \textbf{Client 3} & \textbf{Client 4} \\
\midrule
\multicolumn{5}{l}{\textit{Proportion of Subgroup in Client}} \\
\quad White, Male     & 16.1\% & 44.3\% & 36.8\% & 58.9\% \\
\quad White, Female   & 39.7\% & 14.9\% & 35.1\% & 16.2\% \\
\quad Black, Male     & 16.4\% & 11.9\% & 9.9\%  & 6.4\% \\
\quad Black, Female   & 8.0\% & 16.3\% & 11.4\%  & 6.5\% \\
\quad Asian, Male     & 9.6\% & 2.4\% & 2.1\% & 1.9\% \\
\quad Asian, Female   & 3.5\% & 3.2\% & 0.8\% & 1.2\%  \\
\quad Hispanic, Male     & 4.4\% & 2.1\%  & 3.0\%  & 6.8\% \\
\quad Hispanic, Female   & 2.3\% & 4.9\%  & 0.9\%  & 2.1\% \\
\midrule
\multicolumn{5}{l}{\textit{Outcome Prevalence within Subgroup}} \\
\quad White, Male     & 9.9\% & 19.9\% & 16.6\% & 10.4\% \\
\quad White, Female   & 6.8\% & 3.9\% & 10.0\% & 9.0\% \\
\quad Black, Male     & 10.5\% & 7.8\% & 5.9\% & 10.6\% \\
\quad Black, Female   & 5.2\% & 4.6\% & 10.8\% & 2.8\% \\
\quad Asian, Male     & 15.1\% & 17.1\% & 14.0\% & 4.9\% \\
\quad Asian, Female   & 6.0\% & 3.6\% & 5.9\% & 7.7\% \\
\quad Hispanic, Male     & 7.8\% & 10.8\% & 19.4\% & 10.1\% \\
\quad Hispanic, Female   & 6.1\% & 8.2\% & 0.0\% & 2.1\% \\
\bottomrule
\end{tabular}
\end{table}

\newpage

\subsection{More Experimental Results}\label{supp_fig}

Table \ref{tab:finetuning} shows the values of fine-tuned $\lambda$ and $\gamma$ under different experiment settings.

\begin{table}[H]
\centering
\caption{Fine-tuning configurations for FedIDA under different experimental settings.}
\label{tab:finetuning}
\begin{tabular}{llccc}
\toprule
\textbf{Experiment} & \textbf{Model} & $\boldsymbol{\lambda}$ & \textbf{$\boldsymbol{\gamma}$ fine-tuning range} & $\boldsymbol{\gamma}$ \\
\midrule
\multirow{2}{*}{Adult LR} 
  & FedIDA (FedAvg)     & 2.0 & [0.0112, 0.0223] & 0.0186 \\
  & FedIDA (PFedAvg) & 1.5 & [0.0112, 0.0223] & 0.0149 \\
\midrule
\multirow{2}{*}{Adult FCNN} 
  & FedIDA (FedAvg)     & 2.0 & [0.0112, 0.0223] & 0.0198 \\
  & FedIDA (PFedAvg) & 2.5 & [0.0001, 0.0112] & 0.006266 \\
\midrule
\multirow{2}{*}{ROC LR} 
  & FedIDA (FedAvg)     & 3.0 & [0.0112, 0.0223] & 0.01366 \\
  & FedIDA (PFedAvg) & 2.0 & [0.0223, 0.0334] & 0.02674 \\
\midrule
\multirow{2}{*}{ROC FCNN} 
  & FedIDA (FedAvg)  & 2.5 & [0.0001, 0.0112] & 0.0038 \\
  & FedIDA (PFedAvg) & 3.0 & [0.0112, 0.0223] & 0.01243 \\
\bottomrule
\end{tabular}
\end{table}

Figures S1-S4 present the comparison of predictive and fairness metrics on four different cases per client. Figures 1 and 2 show results on Adult dataset, homogeneously partitioned into five clients. Figures 3 and 4 show results on ROC  dataset, heterogeneously partitioned by age and race into four clients. 
We compare model performance by AUROC, DPD, DPR, DFPR, and DPPV, with error bars representing the 95\% confidence interval of the test metric.

Across clients, FedIDA usually achieves tighter confidence intervals and lower variability in fairness metrics, indicating more stable and equitable model behavior. While standard FL methods often underperform in fairness compared to the centralized model, FedIDA closes this gap and in many cases outperforms both centralized and local models in terms of fairness. These results highlight FedIDA’s ability to enhance fairness without sacrificing predictive performance, making it a compelling solution for equitable federated learning.

\begin{figure}[!htbp]
    \centering
    \includegraphics[scale=0.3]{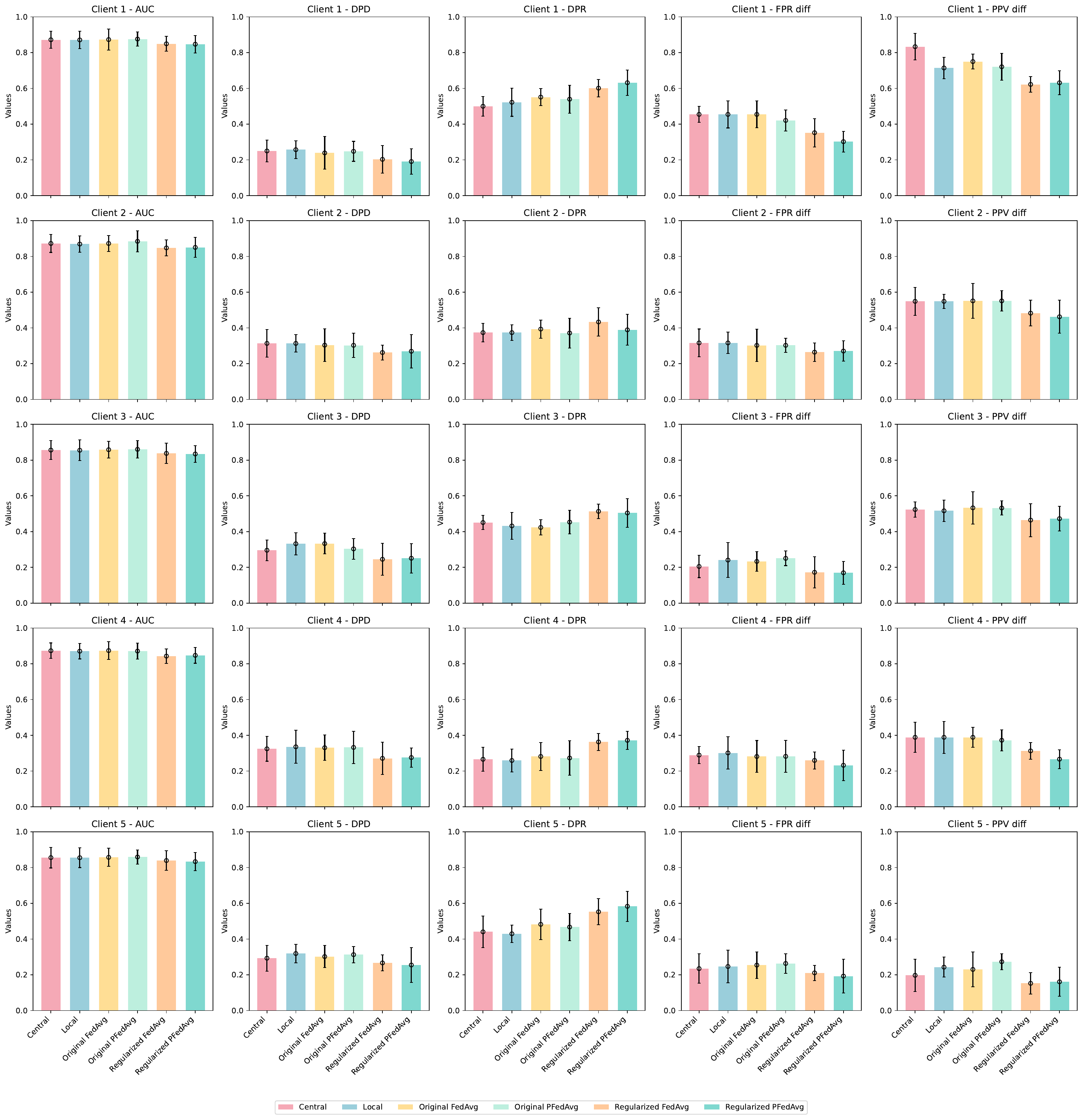}
    \caption{Results for Logistic Regression on Adult dataset}
    \label{fig:adult-lr}
\end{figure}

\begin{figure}[!htbp]
    \centering
    \includegraphics[scale=0.3]{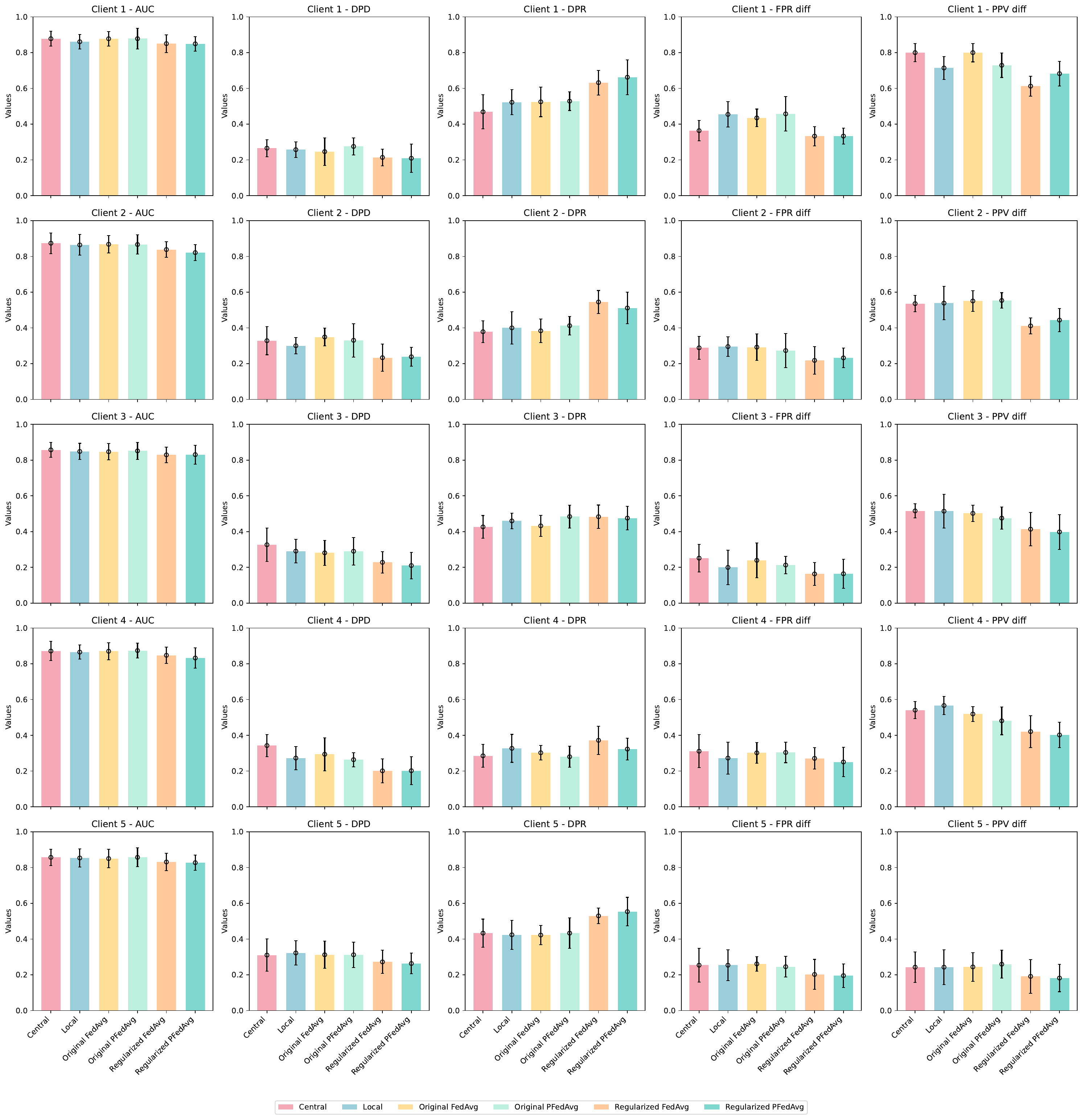}
    \caption{Results for Fully Connected Neural Network on Adult dataset}
    \label{fig:adult-fcnn}
\end{figure}

\begin{figure}[!htbp]
    \centering
    \includegraphics[scale=0.3]{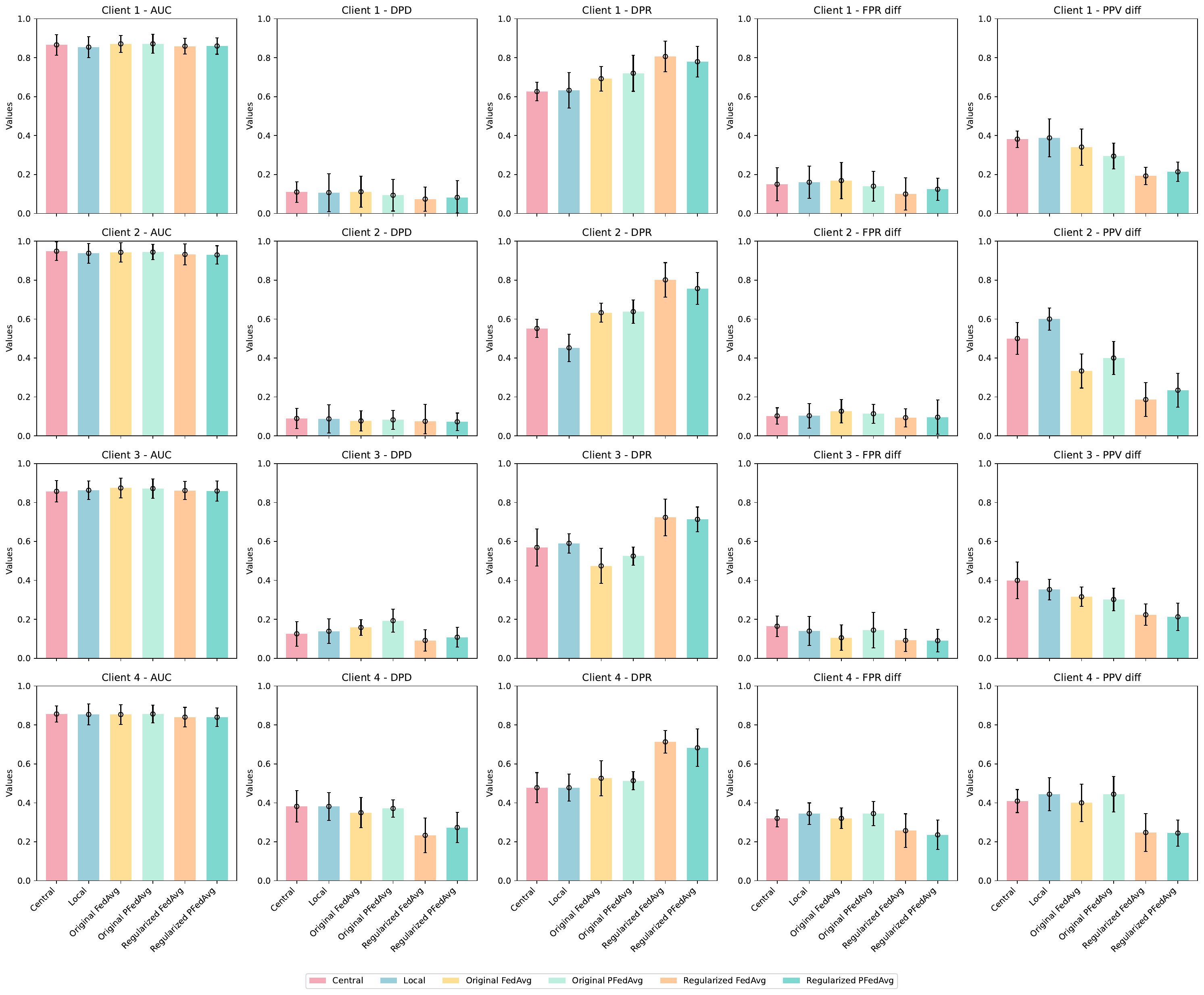}
    \caption{Results for Logistic Regression on ROC dataset}
    \label{fig:roc-lr}
\end{figure}

\begin{figure}[!htbp]
    \centering
    \includegraphics[scale=0.3]{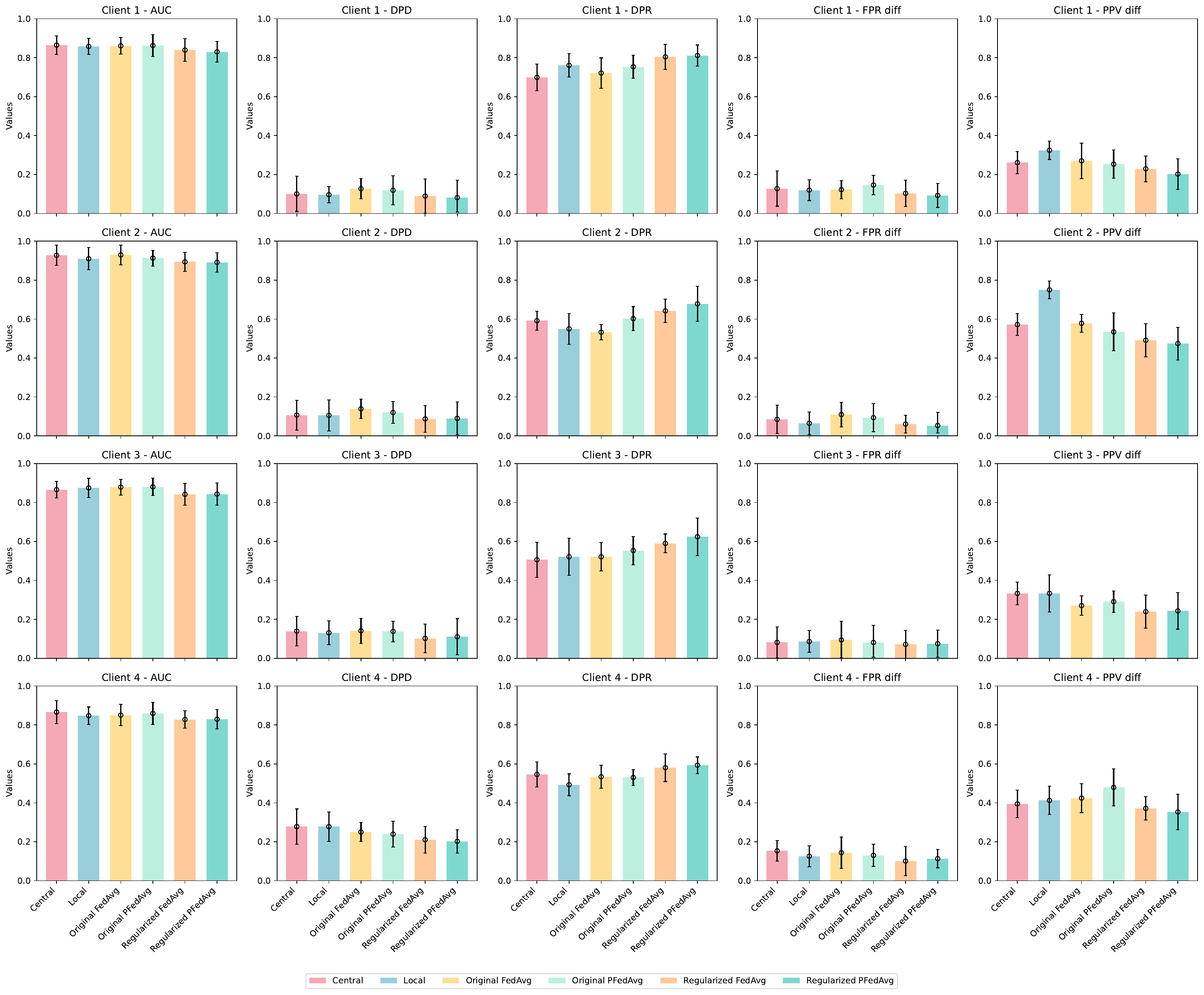}
    \caption{Results for Fully Connected Neural Network on ROC dataset}
    \label{fig:roc-fcnn}
\end{figure}

\FloatBarrier

\end{document}